# Metallic nanoparticles-decorated $Nd_xY_{1-x}Al_3(BO_3)_4$ sub-micrometric particles to enhance anti-Stokes excitation performance

Eloísa G. Hilário[1], Tatiana Habib[1,2], Célio V. T. Maciel[3], Rodrigo F. da Silva[3], Daniel F. Luz[3], Gabriela S. Soares[4], Bruno Caillier[2], Carlos Jacinto,[3] Lauro J. Q. Maia[5], José Maurício A. Caiut[1], and André L. Moura[3,*]

[1]NanoLum – Grupo de Materiais e Sistemas Luminescentes – Departamento de Química – FFCLRP, 14040-901, Ribeirão Preto, São Paulo, Brazil

[2]Laboratoire Diagnostics des Plasmas Hors Equilibre (DPHE), Université de Toulouse, INU Champollion, Albi, France

[3]Programa de Pós-graduação em Física, Instituto de Física, Universidade Federal de Alagoas, Maceió, AL, Brazil

[4]Grupo de Física da Matéria Condensada, Núcleo de Ciências Exatas – NCEx, Campus Arapiraca, Universidade Federal de Alagoas, Arapiraca-AL, Brazil

[5]Grupo Física de Materiais, Instituto de Física, Universidade Federal de Goiás, 74001-970, Goiânia, Goiás, Brazil

*andre.moura@fis.ufal.br

**Abstract**

In the anti-Stokes excitation of trivalent rare-earth ions ($RE^{3+}$), the excitation photons energy is smaller than that of a given absorption transition, and the energy mismatch can be compensated by phonons annihilation from the host lattice. Since the phonon occupation number increases with temperature, heating the system generally increases the efficiency of anti-Stokes excitation. Here, we exploited the intrinsic heating associated with light-to-heat conversion in the interaction of excitation laser light with metallic nanoparticles (Ag or Au) on the surface of submicrometric particles of $Nd_xY_{1.00-x}Al_3(BO_3)_4$ ($x$ = 0.10, 0.20, and 1.00) in order to enhance the efficiency of the anti-Stokes excitation at 1064 nm. Several upconversion emissions are observed from 600 nm to 880 nm, the most intense being at 750 nm due to the $Nd^{3+}$ transition $\{^4F_{7/2}, ^4S_{3/2}\} \rightarrow {}^4I_{9/2}$. Giant enhancements are demonstrated, when compared to undecorated $Nd_xY_{1.00-x}Al_3(BO_3)_4$ particles. The present results can be expanded to other luminescent materials as well as excitation wavelengths.

**1. Introduction**

Nonresonant excitation of trivalent rare-earth ions ($RE^{3+}$) is feasible with phonon creation (Stokes excitation, SE) or annihilation (anti-Stokes excitation, ASE) in order to compensate the energy mismatches [1–4]. In the former, the temperature of the host medium increases, while in the second case cooling can be observed [5]. Of interest here, ASE can provide light emission with photon energy larger than the excitation photon energy as well as emission photons with lower energy. Under ASE, the temperature (T) is a fundamental ingredient, since the phonon occupation number depends on T, and the nonresonant absorption cross-section is given as

$$\sigma_{GSA}(T) = \sigma^0_{GSA}[\exp(\hbar\omega/k_BT)-1]^{-p}, \qquad (1)$$

where $\sigma^0_{GSA}$ is the resonant absorption cross-section, $\hbar\omega$ is the effective phonon energy, $k_B$ the Boltzmann constant, T the host medium temperature, and *p* is the number of phonons participating in the absorption process [6]. Then, heating the system can result in greater light absorption and, consequently, more light emission. The heating can be by intrinsic nonradiative relaxations in the relaxation pathways from the $RE^{3+}$ or extrinsically by use of an external heating source. Other temperature-dependent processes are multiphonon relaxations, and thermal excitation. In the first case, the decay rate is given by

$$W_i^{nr}(T) = W_i^{nr}(T_0)\{[1 - \exp(-\hbar\omega/k_BT)]/[1 - \exp(-\hbar\omega/k_BT_0)]\}^{-p_i}, \qquad (2)$$

in which $p_i$ is the effective phonons number required for the ion to relax to the nearest lower energy level [1]. In the second, $\Lambda_{ij}(T) = C_{ij}P_{ij}(T)$, where $C_{ij}$ is proportional to the electron–phonon coupling strength, and $P_{ij}(T) = [\exp(\hbar\omega/k_BT) - 1]^{-q_{ij}}$ are the phonon occupation numbers, where $q_{ij}$ are the effective phonons number with energy $\hbar\omega$ required to excite the level j from level i [7]. Thanks to the richness of energy levels of some $RE^{3+}$, increasing temperature can trigger these temperature-dependent processes broadening the emission spectrum with light emission at several wavelengths or even decreasing the light-to-light conversion efficiency due to the competition between radiative and nonradiative relaxations. Applications of ASE include harvesting infrared light for solar cells [8], biological fluorescence imaging [9,10], development of lasers [11] and three-dimensional displays [12].

Several authors have been demonstrating ASE of $RE^{3+}$. For example, Thulium ions ($Tm^{3+}$), in different host and architectures, under excitation at around 1064 nm provide light emission at multi-wavelengths from 450 nm to 840 nm, due to ASE followed by multiphonon relaxations and nonresonant excited-state absorption (ESA) [13–20]. Besides the nonresonant excitation with the participation of phonons to compensate the energy mismatches, there is the possibility of multiphoton excitation. The nonresonant ground-state excitation at 1064 nm of Neodymium ions ($Nd^{3+}$) in $Ga_{10}Ge_{25}S_{65}Ga_{10}Ge_{25}S_{65}$ glass occurred due to simultaneous two photon absorption ($^4I_{9/2} \to$ $^4G_{7/2}$) and multiphonon relaxation ($^4G_{7/2} \to$ $^4G_{5/2}$) by isolate $Nd^{3+}$. Also, one-photon ground-state absorption (GSA) with phonon annihilation ($^4I_{9/2} \to$ $^4F_{3/2}$) as well as resonant ESA ($^4I_{11/2} \to$ $^4F_{3/2}$) followed by energy transfer upconversion among $Nd^{3+}$-$Nd^{3+}$ pairs [$^4F_{3/2}$, $^4F_{3/2}$] $\to$ [$^4G_{7/2}$, $^4I_{13/2}$] resulted in light emission at ≈ 535 nm ($^4G_{7/2} \to$ $^4I_{9/2}$), ≈ 600 nm ($^4G_{7/2} \to$ $^4I_{11/2}$; $^4G_{5/2} \to$ $^4I_{9/2}$), and ≈ 670 nm ($^4G_{7/2} \to$ $^4I_{13/2}$; $^4G_{5/2} \to$ $^4I_{11/2}$) [21]. Another remarkable excitation route for getting UC emission is by resonant GSA, followed by thermal-excitation from an excited state to the closest upperlying level. Giant enhancement (670-fold) of $Nd^{3+}$ emission centered at 754 nm ({$^4F_{7/2}$, $^4S_{3/2}$} $\to$ $^4I_{9/2}$), when increasing the glass temperature (200 K to 535 K), was observed exciting resonantly the $Nd^{3+}$ ($^4I_{9/2} \to$ {$^4F_{5/2}$, $^2H_{9/2}$}) at 805 nm {$^4F_{5/2}$, $^2H_{9/2}$} $\to$ {$^4F_{7/2}$, $^4S_{3/2}$} [22]. Similar excitation pathway due to resonant ESA followed by thermal excitation to upperlying levels were reported [23,24]. Remarkably, ladder-thermal excitation involving more than 2 levels was observed under resonant excitation at 866 nm ($^4I_{9/2} \to$ $^4F_{3/2}$) of $Nd^{3+}$ in a fluoridate glass [23], and under nonresonant excitation (1060 nm) of $Nd^{3+}$ in $LaPO_4$ nanocrystals [25]. The population of the {$^4F_{7/2}$, $^4S_{3/2}$} states occurred after the following sequence of thermal-excitation $^4F_{3/2} \to$ {$^4F_{5/2}$, $^2H_{9/2}$} $\to$ {$^4F_{7/2}$, $^4S_{3/2}$} resulting in light emission at around 750 nm ({$^4F_{7/2}$, $^4S_{3/2}$} $\to$ $^4I_{9/2}$) [23,25].

The photon-avalanche (PA) is a fascinating excitation mechanism for $RE^{3+}$. Some requirements for the PA are the GSA cross-section ($\sigma_{GSA}$) much smaller than an ESA cross-section ($\sigma_{ESA}$) with $\sigma_{GSA}/\sigma_{ESA} < 10^{-4}$ [26]. Under these circumstances, the material is initially transparent to the excitation laser wavelength ($\lambda_{exc}$). However, with low probability, an excited state can be populated (B), for example with phonon creation (Fig. 1). Considering the $\lambda_{exc}$ is resonant with the ESA B $\to$ C transition, and the occurrence of energy transfer (cross-relaxation) between an ion at C and another at the ground state (A), both ions are promoted to the intermediate state (B). Given the resonance of $\lambda_{exc}$ with the ESA transition B $\to$ C, both ions can be promoted to the excited state C. In resume, after the sequence of events of a weakly GSA, resonant ESA, cross-relaxation, and ESA, there are two ions at the state C (Fig. 1). The repetition of this sequence of events can double the population in the state C at every iteration, leading to giant enhancements of the excitation photons absorption and photoluminescence (PL). Some signatures of PA are a threshold ($P_{th}$) behavior in the PL input-output power dependence, which presents a *S*-shape with saturation at large excitation powers ($P_{exc}$); enlargement of the PL dynamics for $P_{exc}$ close to $P_{th}$; and shortening of the dynamics upon increasing $P_{exc}$ beyond $P_{th}$. It is important to point out that, most of the PA excitation of $RE^{3+}$ were demonstrated at low temperatures [27–31]. PA excitation of $Nd^{3+}$ at room [32–34] or higher temperature [35] was already demonstrated. In the last case, Marciniak et al. [35] excited at 1064 nm $Nd^{3+}$ ions in different nanocrystals in the form of powders (Y2O3, Gd2O3, YGdO3, YAlO3, Y3Al5O12, and LiLaP4O12). A P$_{th}$ was observed in the input-output PL, but without the S-shape characteristic. The PL dynamics was determined using theoretical model based on coupled-rate equations. Thanks to the large population redistribution, the authors also investigated the suitability of these systems for ratiometric optical thermometry. In a subsequent work by da Silva et al. [36], a PA-like mechanism was demonstrated in NdAl$_3$(BO$_3$)$_4$ submicrometric particles under excitation at 1064 nm. In the PA-like process, the heat generation by phonon emissions increased the particles temperature, which was a fundamental tool to trigger the mechanism. Accordingly, the same group demonstrated the triggering of the PA-like by heating the system externally using electric resistive wires [37] and by means of an auxiliary beam at 808 nm, which switched on the avalanche by increasing the rate of phonon emissions [38].

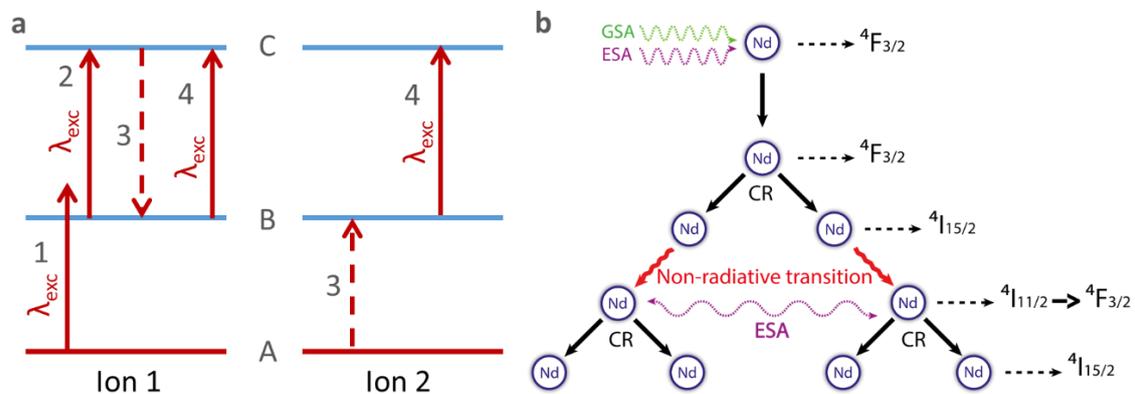

**Fig. 1.** (a) Representation of the photon avalanche mechanism illustrating the sequence of events that starts with one excited ion at state B and ends with two ions at the excited state C. 1, 2, 3, and 4 stand for nonresonant ground-state excitation, resonant excited-state absorption, cross-relaxation, and resonant excited-state absorption, respectively. (b) Illustration of the photon-avalanche like mechanism for Nd$^{3+}$ excited at 1064 nm. CR, and ESA stand for cross-relaxation ($^4F_{3/2}$, $^4I_{9/2}$ → $^4I_{15/2}$, $^4I_{15/2}$) and resonant excited-state absorption ($^4I_{11/2}$ → $^4F_{3/2}$), respectively.

Here, we decorate the crystalline Nd$_x$YAl$_{1.00-x}$(BO$_3$)$_4$ particles with metallic nanoparticles (MNPs), Au or Ag, in order to investigate their influence on the ASE at 1064 nm. It is well-known that MNPs can enhance the PL of RE$^{3+}$ by increasing the local field acting on ions located in close proximity to them, as well as by energy transfer from the MNPs to the RE$^{3+}$ [39–45]. However, we demonstrate that large light-to-heat conversion, associated with the interaction of the excitation laser with the MNPs, increases the crystalline particles temperature, which is a fundamental ingredient for the ASE in Nd$^{3+}$ at 1064 nm. Given the coexistence of several excitation and emissions processes from the Nd$^{3+}$, light emission due to several transitions were observed whose dynamics depend on *x* as well as the presence of Au or Ag nanoparticles.

## 2. Experimental Procedure
### 2.1. Synthesis of Metallic nanoparticles-decorated Y$_{1-x}$Nd$_x$Al$_3$(BO$_3$)$_4$

The $Y_{1-x}Nd_xAl_3(BO_3)_4$ powders were prepared using the Pechini method. Stoichiometric amounts of $Y(NO_3)_3$ at 0.48 M and $Nd(NO_3)_3$ at 0.12 M were added to 50 mL of ultrapure water. The metals were complexed with citric acid in a molar ratio of 3:1 citric acid to metals (including boron). After heating and stirring at 75°C for 30 minutes, $Al(NO_3)_3 \cdot 9H_2O$ diluted in 20 mL of water was added, and it was keeping under heating and stirring for 15 minutes. Then, a solution containing boric acid and D-sorbitol was added at a mass ratio of 3:2 citric acid to D-sorbitol. The temperature was then raised to 100 °C, and the solution was heated for 1 hour to obtain a resin. The resin was dried in an oven overnight at 120 °C. Finally, the material was heat-treated under an $O_2$ atmosphere using a heating stage of 400 °C/24 hours, 700 °C/24 hours, and 1100 °C/5 minutes.

**2.2. metallic particles decoration**

Silver decoration: 0.1 g of $Y_{1-x}Nd_xAl_3(BO_3)_4$ particles were dispersed in 10 mL of aqueous solution composed of $AgNO_3$ ($5 \times 10^{-3}$ M) and trisodium citrate (0.05 M) under stirring. The reaction media was exposed 5 minutes of He plasma jet treatment. The particles were isolated by centrifuging at 14500 rpm/5 minutes. Gold decoration: 0.1 g of $Y_{1-x}Nd_xAl_3(BO_3)_4$ particles were dispersed in 10 mL of aqueous solution composed of $HAuCl_4$ ($0.1 \times 10^{-3}$ M) and PVP 40 - Polyvinylpyrrolidone ($0.05 \times 10^{-3}$ M), under stirring. The reaction media was exposed 5 minutes of He plasma jet treatment. The particles were isolated by centrifuging at 14500 rpm / 5 minutes. The device used to generate the plasma jet consists of a power supply which works at fast voltage pulses on a capacitive load, a Helium gas bottle connected to a gas flowmeter to regulate the gas intake and an asymmetric glass source. Further detail was showed in a previous publication [46].

**2.3. Optical experiment**

The particle powders were placed on a metallic support and pressed gently. The excitation beam at 1064 nm, due to a continuous-wave Nd:YAG laser, was guided to the powder through a hot mirror and focused by a 10 cm focal length lens at normal incidence to the powder surface (Fig. 2). The beam diameter at the powder surface was ≈ 30 µm, determined by the knife edge technique. The PL was collected along this normal line by a telescope composed of two 5 cm focal length lenses and focused onto a multimode optical fiber coupled to a spectrometer equipped with a charge-couple device (CCD). This spectrometer allowed simultaneous spectral measurements from 330 nm to 1180 nm with resolution of 2.0 nm. The CCD integration time was fixed to allow consecutive spectrum acquisition every 50 ms. The hot mirror rejected most of the elastically-scattered laser light.

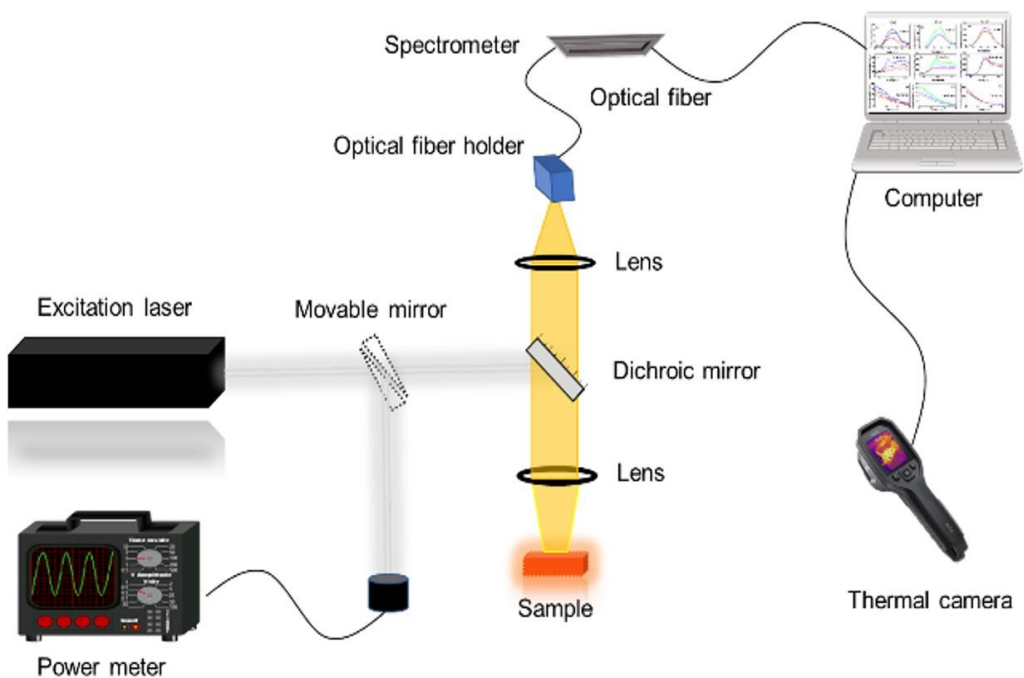

**Fig. 2.** Experimental setup representation for the anti-Stokes excitation of $Nd^{3+}$ at 1064 nm.

The interaction of the laser light with the particles generated heat associated with phonon emission in the relaxation pathways from the $Nd^{3+}$ as well as in the interaction of the excitation light with the MNPs. Then, we used a thermal-camera to measure the powder temperature: FLIR-E40 thermographic camera; detection range, $-20 \leq T \leq 650$ °C; accuracy 2 °C. The maximum temperature at the excited volume was recorded as a function of exposure time of the particles to the excitation laser. Those temperature measurements were made simultaneously with the spectral emission detection allowing the association of each spectrum to the powder temperature.

The experimental procedure to characterize the PL consisted in irradiate the powders with different values of excitation power ($P_{exc}$). In each run, the excitation beam was unblocked while acquiring the PL spectra and monitoring the particles temperature. Noteworthy, in order to avoid irreversible structural modification of the dielectric particles, the maximum $P_{exc}$ was limited to obtain temperatures below 650 °C, which is lower than the maximum temperature used in the heat treatment (1100 °C). The reproducibility of the experimental results was checked by performing consecutive experimental realizations under the same starting conditions.

## 3. Results and Discussion

The unconventional excitation of $Nd^{3+}$ at 1064 nm starts with phonon-assisted transitions $^4I_{9/2}$ → $^4F_{3/2}$ and/or with the thermal excitation of the $^4I_{11/2}$ from the ground state ($^4I_{9/2}$) followed by the resonant ESA $^4I_{11/2} \rightarrow {}^4F_{3/2}$ (Fig. 3). Both processes are temperature-dependent with low probabilities at room temperature. However, once excited at $^4F_{3/2}$, cross-relaxation with another ion at the ground state ($^4I_{9/2}$) transfers both ions to the $^4I_{15/2}$, from which the two ions relax with phonon emission to the lower-lying state $^4I_{13/2}$ and subsequently to the $^4I_{11/2}$. At the $^4I_{11/2}$ state, the resonant ESA ($^4I_{11/2} \rightarrow {}^4F_{3/2}$) leads both ions to the $^4F_{3/2}$ level. Continuing, one can have the establishment of an energy-looping, which magnifies the population of the $^4F_{3/2}$ state and can evolve to an avalanche mechanism (Fig. 1b presents an illustration). The main difference between both processes is the absence of threshold in the energy-looping regime. Besides the emission at 880 nm ($^4F_{3/2} \rightarrow {}^4I_{9/2}$), the nonradiative relaxations with phonon emission provide intrinsic heating of the particles, which favors the phonon-assisted excitation processes as well as provides thermal excitations from the $^4F_{3/2}$ to upperlying states whose radiative relaxations enrich the emission spectrum. The occurrence of energy-looping or PA-like in $Nd_xY_{1.00-x}Al_3(BO_3)_4$ powders depends on $x$ since the cross-relaxation of $Nd^{3+}$-$Nd^{3+}$ pairs depends strongly on the distance of the ions: energy-looping for $x \leq 0.20$, while for $0.40 \leq x \leq 1.00$ the systems evolve to the PA-like mechanism [47]. Here, the powders investigated are, in principle, in the energy-looping ($x = 0.10$, and 0.20) and PA-like ($x = 1.00$) regimes.

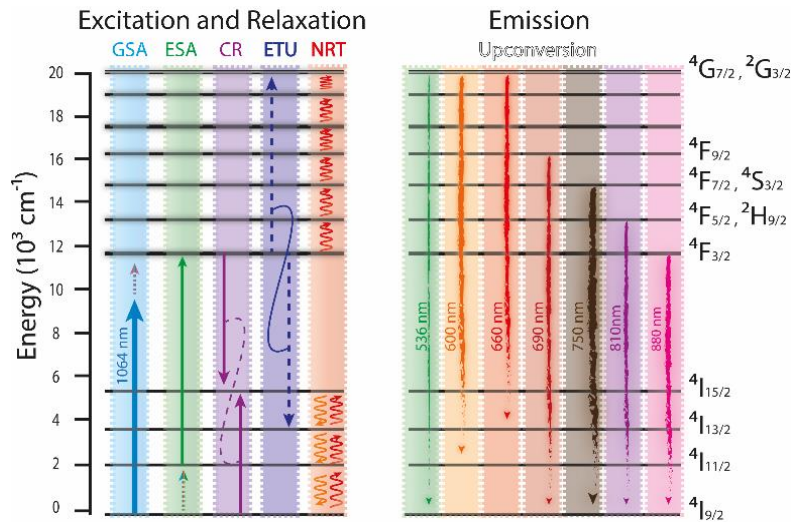

**Fig. 3.** Partial energy level diagram of the $Nd^{3+}$ illustrating the excitation and relaxation pathways for a nonresonant excitation at 1064 nm. GSA, ESA, CR, ETU, and NRT stand for ground-state absorption, excited-state absorption, cross-relaxation, energy-transfer upconversion, and nonradiative transitions involving phonon creation (downward wavy arrow) or annihilation (upward wavy arrow), respectively. The radiative transitions and the corresponding wavelengths are indicated in the figure right side.

All $Y_{1-x}Nd_xAl_3(BO_3)_4$ particles were structurally and morphologically characterized, before and after decoration, by XRD (no show), diffuse reflectance spectra, and electronic microscopy analysis. Figure 4a shows the diffuse reflectance whose several peaks are observed and designed to ground-state absorptions of the $Nd^{3+}$ ions and also exhibits the plasmon resonance of the particles decorated with Au and Ag nanoparticles. Figures 4b-d are transmission electron microscopy (TEM) images of the Ag:Nd$_x$Y$_{1.00-x}$Al$_3$(BO$_3$)$_4$ ($x$ = 0.1) while in Figs. e-g are TEM images of the Au:Nd$_x$Y$_{1.00-x}$Al$_3$(BO$_3$)$_4$ ($x$ = 0.1). Figures 4h and 4i present size histograms of the Au:Nd$_x$Y$_{1.00-x}$Al$_3$(BO$_3$)$_4$ and Ag: Nd$_x$Y$_{1.00-x}$Al$_3$(BO$_3$)$_4$ MNPs, respectively.

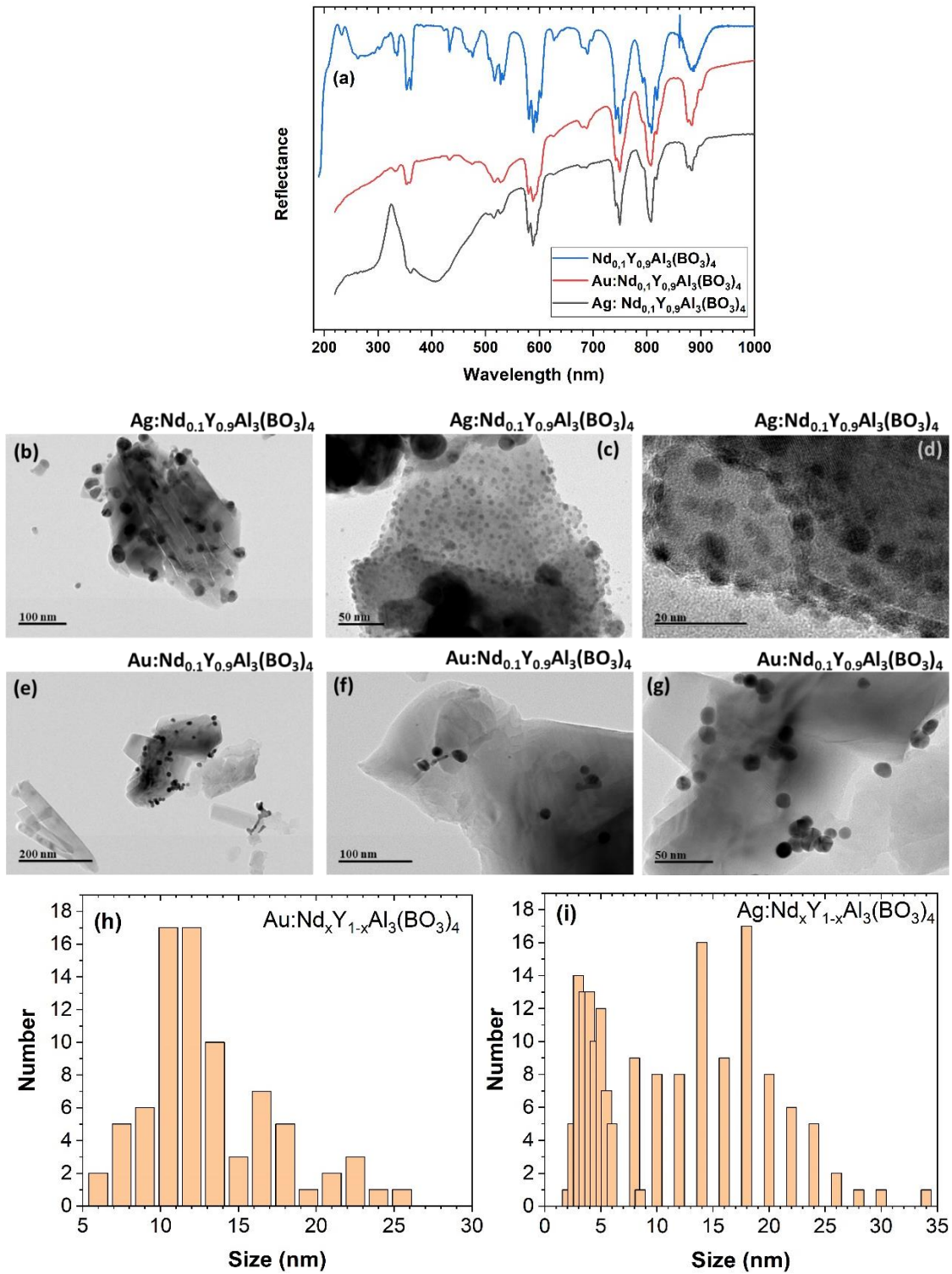

**Fig. 4.** (a) Nd$_x$Y$_{1.00-x}$Al$_3$(BO$_3$)$_4$ ($x$ = 0.1) powder characterization by diffuse reflectance in which several ground-state absorption transition due to the Nd$^{3+}$ are observed, and the presence of plasmons resonance for the particles decorated with Au and Ag nanoparticles. (b-d) TEM of the Ag:Nd$_x$Y$_{1.00-x}$Al$_3$(BO$_3$)$_4$ ($x$ = 0.1). (e-g) TEM of the Au:Nd$_x$Y$_{1.00-x}$Al$_3$(BO$_3$)$_4$ ($x$ = 0.1); and Metallic Particles size distributions to (h) Au:Nd$_x$Y$_{1.00-x}$Al$_3$(BO$_3$)$_4$ and (i) Ag: Nd$_x$Y$_{1.00-x}$Al$_3$(BO$_3$)$_4$.

The PL characterization was performed for several values of P$_{exc}$ (Figs. 5 and 6). For $x$ = 0.10 the PL intensity was very low compared to the noise of the detection system (Fig. 5a). For $x$ = 0.20 (Fig. 5b) the stronger emission is at 880 nm (P$_{880}$, $^4$F$_{3/2}$ → $^4$I$_{9/2}$) whose emitting level ($^4$F$_{3/2}$) is the first populated in the excitation pathways from the Nd$^{3+}$. P$_{880}$ presents monotonous dependence with P$_{exc}$. Notice that, no upconversion emission was detected for $x$ = 0.10, while it was seen easily for $x$ = 0.20 (Figs. 5a and b). In the first case, neither the energy-looping nor the PA-like are relevant. The growth of the emissions without threshold for $x$ = 0.20 means the establishment of an energy-looping thanks to the reduction of the distance between Nd$^{3+}$-Nd$^{3+}$ pairs, which favors the cross-relaxation ([$^4$F$_{3/2}$, $^4$I$_{9/2}$] → [$^4$I$_{15/2}$, $^4$I$_{15/2}$]), a key element for the energy-looping. The particles temperature increases due to the phonon emission (right axes in Fig. 5), favoring the phonon-assisted excitation processes. Based on the emission spectra as a function of the exposure time to the excitation laser (Fig. 6a, for $x$ = 0.20), the excitation dynamics is quite slow (≈ 1 s) due to the cycle: absorption of light leads to temperature increase, which enhances the light absorption. One should notice the absorption processes involve phonon annihilation, while the relaxation pathways from the Nd$^{3+}$ ions generate phonons. The increase of the particles temperature signalizes the prevalence of phonon creation. The population of the upperlying states (above the $^4$F$_{3/2}$) can be by distinct processes. The relevant processes here are the energy-transfer upconversion (ETU) [$^4$F$_{3/2}$, $^4$F$_{3/2}$] → [$^4$I$_{13/2}$, {$^4$G$_{7/2}$, $^3$G$_{3/2}$}]) and ladder-thermal excitations from the metastable state $^4$F$_{3/2}$. Given the maximum temperature attained by the particles (≈ 120 °C for $x$ equals 0.20), the population of the {$^4$F$_{5/2}$, $^2$H$_{9/2}$} (from where, ground-state relaxation provides light at around 810 nm) should be by thermal excitation from the lowerlying state $^4$F$_{3/2}$. For the more energetic emissions, ETU can populate (with low efficiency due to the small content of Nd$^{3+}$) the {$^4$G$_{7/2}$, $^2$G$_{3/2}$} (emission at 600 nm and 660 nm), from which successive nonradiative relaxations with phonon emissions populate the $^4$F$_{9/2}$ (emission at 690 nm), and {$^4$F$_{7/2}$, $^4$S$_{3/2}$} (emission at 750 nm) states. Other contribution can be by ladder-thermal excitation $^4$F$_{3/2}$ → {$^4$F$_{5/2}$, $^2$H$_{9/2}$} → {$^4$F$_{7/2}$, $^4$S$_{3/2}$}, which, as discussed below, should be relevant for the particles decorated with Au and Ag, was well as the dielectric particles with $x$ = 1.00, in which the attained temperate are much larger.

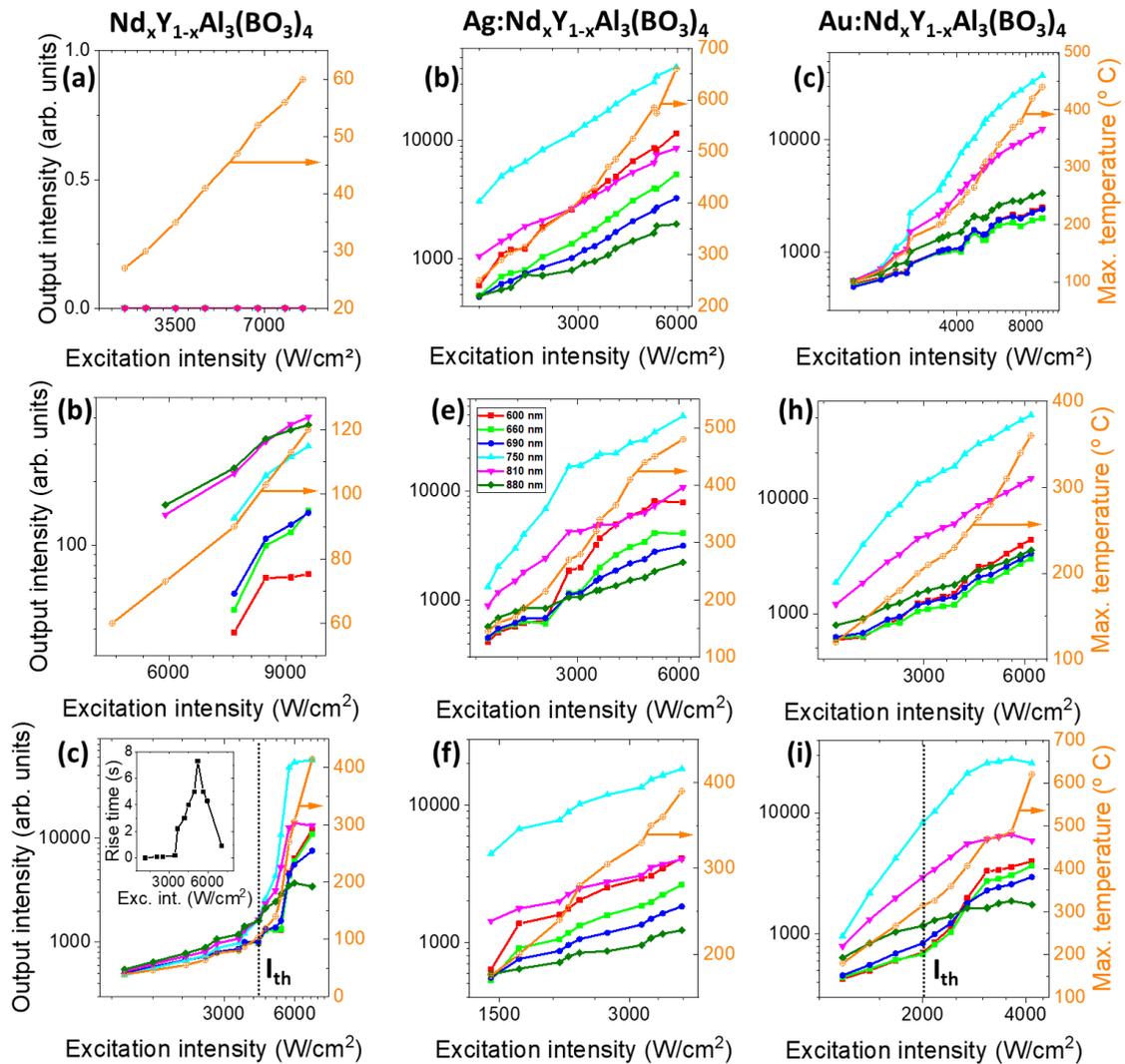

**Fig. 5.** Photoluminescence characterization by the input-output power dependences for the $Nd_xY_{1.00-x}Al_3(BO_3)_4$ powders [$x$ = 0.10 (first row), 0.20 (second row), and 1.00 (third row] without (first column) and with Ag (second column) and Au (third column) metallic nanoparticles on the surfaces of the crystalline particles. The light emission at 880 nm, 810 nm, 750 nm, 690 nm, 660 nm, and 600 nm are associated with $Nd^{3+}$ transitions $^4F_{3/2} \rightarrow {^4I_{9/2}}$, $\{^4F_{5/2}, {^2H_{9/2}}\} \rightarrow {^4I_{9/2}}$, $\{^4F_{7/2}, {^4S_{3/2}}\} \rightarrow {^4I_{9/2}}$, $^4F_{9/2} \rightarrow {^4I_{9/2}}$, $\{^4G_{7/2}, {^2G_{3/2}}\} \rightarrow {^4I_{13/2}}$, and $\{^4G_{7/2}, {^2G_{3/2}}\} \rightarrow {^4I_{13/2}}$, respectively. The inset in Fig. 5c represents the build-up time of the photoluminescence, measured as the stationary intensity times [1 - exp(-1)].

For the particles with $x$ = 1.00, the PL differs from $x$ = 0.20 markedly with the presence of an excitation power threshold ($P_{th} \approx 1.5$ W, corresponding to an excitation intensity threshold $I_{th} \approx 5000$ W/cm²) from which the PL increases by several times (Fig. 5c), and by the dramatic change in the PL spectra (Fig. 6b and c for $P_{exc}$ lower and larger than $P_{th}$, respectively). The inset of Fig. 5c shows the build-up time of the PL ($x$ = 1.00), measured as the stationary intensity times [1 - exp(-1)]. It is evident the slowdown of the PL dynamics for $P_{exc} \approx P_{th}$, and its shortening for $P_{exc} > P_{th}$, one characteristic of the PA process. Accordingly, the particles temperature kinks at $P_{exc} = P_{th}$ indicating the enhancement of both excitation light absorption and the phonon emission. Although the $^4F_{3/2}$ state is the first populated in the excitation pathway, the emission at 880 nm presents a small growth followed by saturation, i.e., it displays the S-shape characteristic curve of a PA. It is worth emphasizing that we are calling PA-like because the participation of phonons, which increases the particles temperature favoring the several thermal-assisted processes [36]. The small growth of the emission at 880 nm is associated to population losses to the upperlying levels by thermal excitation thanks to the large attained temperatures. The ladder-thermal excitation from the $^4F_{3/2}$ state populates the {$^4F_{5/2}$, $^2H_{9/2}$} and {$^4F_{7/2}$, $^4S_{3/2}$} levels, whose ground-state relaxations radiate at 810 nm and 750 nm, respectively. Consistently, when increasing P$_{exc}$, the emission at 810 nm saturates first, while that one at 750 nm presents the largest enhancement (≈ 20-fold). With the population growth of the {$^4F_{7/2}$, $^4S_{3/2}$} levels and the high temperature of the particles under large P$_{exc}$, intense emissions emerge at 690 nm ($^4F_{9/2} \rightarrow {}^4I_{9/2}$), 660 nm ({$^4G_{7/2}$, $^2G_{3/2}$} $\rightarrow {}^4I_{13/2}$), and 600 nm ({$^4G_{7/2}$, $^2G_{3/2}$} $\rightarrow {}^4I_{11/2}$). The emission at 690 nm shows a trend of saturation with P$_{exc}$, while those ones at 660 nm and 600 nm, whose electronic transitions leave the same state, continue to increase, signaling that the ladder-thermal excitation {$^4F_{7/2}$, $^4S_{3/2}$} $\rightarrow {}^4F_{9/2} \rightarrow$ {$^4G_{7/2}$, $^2G_{3/2}$} is the main mechanism for populating issuing states.

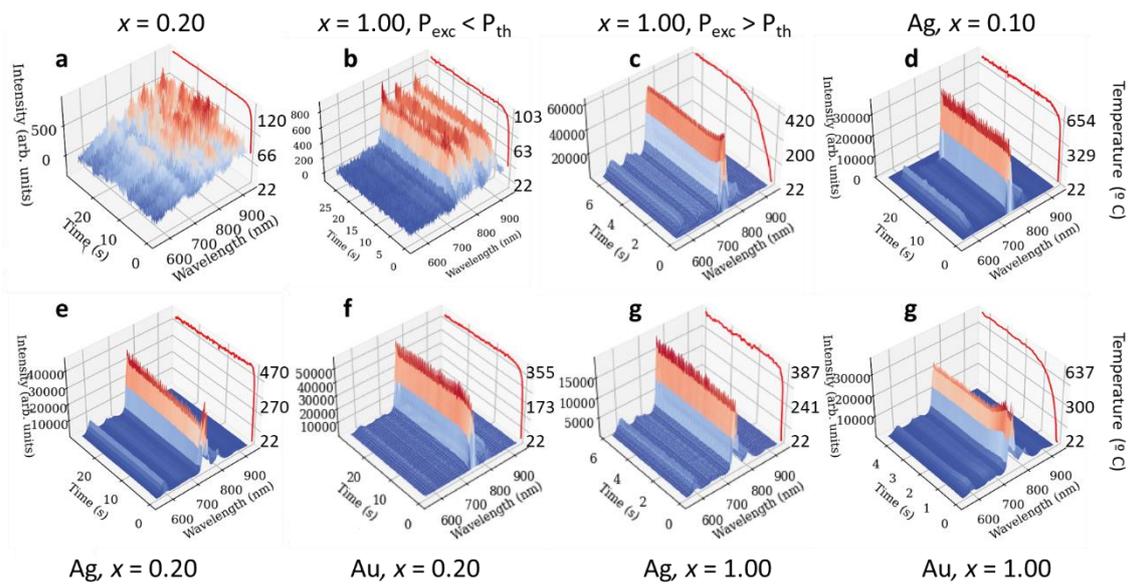

**Fig. 6.** Representative spectra of undecorated and decorated (Ag and Au) Nd$_x$Y$_{1.00-x}$Al$_3$(BO$_3$)$_4$ particles. The spectra are given as a function of the exposure time to the excitation laser. The particles temperature, which increased intrinsically due to the light-to-heat conversion, are also represented.

In the next step, we investigated the effect of decorating the crystalline Nd$_x$Y$_{1.00-x}$Al$_3$(BO$_3$)$_4$ particles with MNPs (Ag and Au). It is well-known that plasmon states, associated with collective oscillations of electrons at the metal-dielectric interfaces, can affect the radiative transition rates of RE$^{3+}$ [39–42,48,49]. The surface plasmon resonance needs to overlap with some specific absorption or emission transition band in order to modify the PL characteristics significantly. The superposition of the plasmon resonance with the $Nd^{3+}$ emissions is weak at the wavelengths of interest here, i.e., 600 nm, 660 nm, and 690 nm, 750 nm, 810 nm, and 880 nm for Au and Ag decorated $Nd_xY_{1.00-x}Al_3(BO_3)_4$ particles, and there is no superposition with the excitation wavelength (1064 nm) (Fig. 4a). Since the interaction of the plasmon field with the $RE^{3+}$ depends on the distance between the MNPs and the $RE^{3+}$ [50], and the MNPs are placed on the surface of the $Nd_xY_{1.00-x}Al_3(BO_3)_4$ dielectric particles, enhancement of PL properties associate with plasmons is not expected to be significant. On the other hand, there is the light-to-heat conversion in the interaction of the electromagnetic field with the MNPs due to the absorption of excitation light by vibrational transitions of the MNPs [43,44,51,52]. As a consequence, it can increase the $Nd_xY_{1.00-x}Al_3(BO_3)_4$ dielectric particles temperature, affecting the several temperature-dependent processes.

Pristine $YAl_3(BO_3)_4$ particles without $Nd^{3+}$ and MNPs were excited at 1064 nm under $P_{exc}$ = 0.5 W (corresponding intensity $I_{exc}$ = 1666 W/cm$^2$). Low temperature rise was observed (< 40 °C, Fig. 7a), which is attributed to multiphoton absorption of the $YAl_3(BO_3)_4$ bandgap and surface defects followed by nonradiative relaxations. For *x* = 0.20 and 1.00, the nonradiative relaxations from $Nd^{3+}$ plays a significant role (Figs. 5b and c), since the observed temperatures are larger than for undoped $YAl_3(BO_3)_4$ particles. Notably, a greater temperature increase was unveiled when exciting, under the same experimental conditions, the undoped $YAl_3(BO_3)_4$ particles decorated with Au and Ag (Fig. 7a) owing to the light-to-heat conversion in the interaction of the excitation laser with the MNPs. In addition, the Ag decorated $YAl_3(BO_3)_4$ particles showed a higher temperature than those decorated with Au, which is a consequence of the higher concentration of Ag NP on the dielectric particles surface (Fig 4 b-g). For $Nd_xY_{1.00-x}Al_3(BO_3)_4$ particles decorated with Ag and Au, there are two relevant sources of heat: nonradiative relaxations from the $Nd^{3+}$ (*x* = 0.20, and 1.00), and excitation of vibrational modes of the MNPs (*x* =0.10, 0.20, and 1.00). Then, the $Nd_xY_{1.00-x}Al_3(BO_3)_4$ (*x* = 0.10, 0.20, and 1.00) particles decorated with both Ag and Au nanoparticles showed large temperature rise (second and third columns of Fig. 5, respectively) when compared with the corresponding undecorated dielectric particles (first column of Fig. 5), impacting directly on the PL (Figs. 5, 6 and 7). As explained above, the anti-Stokes excitation of the $Nd^{3+}$ at 1064 nm starts with two temperature-dependent processes: nonresonant GSA ($^4I_{9/2} \rightarrow {}^4F_{3/2}$) and resonant ESA ($^4I_{11/2} \rightarrow {}^4F_{3/2}$) with the population of the $^4I_{11/2}$ state by thermal coupling with the ground state ($^4I_{9/2}$). Consequently, the additional temperature rise promoted by the interaction of the excitation beam with the MNPs favors both excitation pathways. Accordingly, the PL intensity of the observed $Nd^{3+}$ transitions are greater for Ag and Au decorated $Nd_xY_{1.00-x}Al_3(BO_3)_4$ particles than undecorated ones (Fig. 5 and Fig. 7).

For x = 0.10 and 0.20, giant enhancements are revealed for the emissions at 750 nm ({$^4F_{7/2}$, $^4S_{3/2}$} → $^4I_{9/2}$), and 810 nm ({$^4F_{5/2}$, $^2H_{9/2}$} → $^4I_{9/2}$), respectively. Similar to the undecorated particles, the $^4F_{3/2}$ state is the first populated in the excitation pathways from the $Nd^{3+}$, but the corresponding PL at 880 nm is quenched by thermal excitations to upperlying states. In this sense, for powders with Ag and Au, the emission at 880 nm presented small enhancements. In the sequence of thermal excitations, $^4F_{3/2}$ → {$^4F_{5/2}$, $^2H_{9/2}$} → {$^4F_{7/2}$, $^4S_{3/2}$}, a large population is attained in the {$^4F_{7/2}$, $^4S_{3/2}$} states, which is inferred by the high relative intensities (Fig. 5, and Figs. 7b-d). It is worth noticing that for undecorated particles (*x* = 0.20), the population of the {$^4F_{7/2}$, $^4S_{3/2}$} states presents contributions from ladder-thermal excitation from the lowerlying states. For the most energetic levels, {$^4G_{7/2}$, $^2G_{3/2}$} and $^4F_{9/2}$, ETU dominates over ladder-thermal excitations (undecorated *x* = 0.20), and the emissions whose $Nd^{3+}$ electronic transitions departure from those states, 600 nm, 660 nm, and 690 nm present small intensities in the exploited range of $P_{exc}$ (Fig. 5b). However, the emission at those wavelengths are pronouncedly larger for Ag than Au decorated particles, which is explained by the higher temperature attained by Ag decorated particles in the same range of $P_{exc}$ (second and third column of Fig. 5). Based on that, in the excitation pathway of the $Nd^{3+}$ {$^4G_{7/2}$, $^2G_{3/2}$} and $^4F_{9/2}$ states (*x* = 0.10 and 0.20), the ladder-thermal excitation should dominate over ETU for Ag and Au decorated particles thanks to the large attained temperatures.

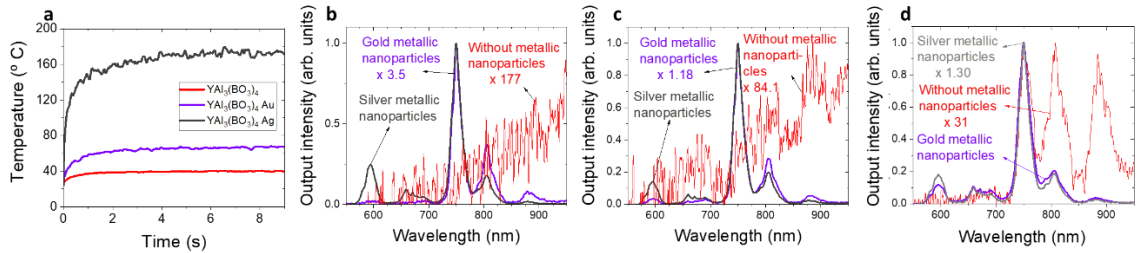

**Fig. 7.** (a) Temperature rise of the $Nd^{3+}$ undoped (*x* = 0.00) $Nd_xY_{1.00-x}Al_3(BO_3)_4$ particles undecorated and decorated with Ag and Au nanoparticles under excitation at 1064 nm ($P_{exc}$ = 0.5 W, $I_{exc}$ = 1666 W/cm²). Normalized photoluminescence spectra of the $Nd_xY_{1.00-x}Al_3(BO_3)_4$ particles without and with Ag and Au nanoparticles for *x* equals (b) 0.10, (c) 0.20, and (d) 1.00 obtained using the excitation powers ($P_{exc}$) of 1.7 W ($I_{exc}$ = 5666 W/cm²), 1.7 W ($I_{exc}$ = 5666 W/cm²), and 1.2 W ($I_{exc}$ = 4000 W/cm²), respectively.

Peculiarly, the sample with *x* = 1.00 displays the PA-like behavior when undecorated with MNPs, which was suppressed for Ag decorated particles (Fig. 5f). This is attributed to the thermalization at high temperatures, which increases the thermal coupling $^4I_{9/2}$ → $^4I_{11/2}$. As a consequence, the 1064 nm beam would present high excitation rates to the $^4F_{3/2}$ via the ESA $^4I_{11/2}$ → $^4F_{3/2}$ transition, and the threshold behavior would be suppressed because the PA requirement of weak ground-state absorption followed by cross-relaxation to populate an intermediate state ($^4I_{11/2}$) is not fulfilled. Remarkably, the *x* = 1.00 powder decorated with Au nanoparticles presented a mixing of thermal-assisted absorption enhancement and PA-like evidenced by the presence of a threshold at $P_{exc}$ = 0.80 W (corresponding to $I_{exc}$ = 2666 W/cm$^2$) (Fig. 5i), from which the emission at 750 nm saturated while the most energetic at 600 nm, 660 nm, and 690 nm increased abruptly. This threshold is lower than that observed for undecorated $NdAl_3(BO_3)_4$ particles, ≈ 1.5 W (5000 W/cm$^2$), whose decrease is consistent with previous results, which showed a decrease of the PA-like threshold when increasing, by external means, the starting particles temperature (20 < T < 215 °C) [37].

At this point, it should be evident that the decoration of dielectric particles with MNPs benefits the ASE due to the light-to-heat conversion, which rises the dielectric particles temperature favoring the ASE with the enhancement of the phonon occupation number. As last remarks, we mentioned that despite the large phonon energy of $Nd_xY_{1.00-x}Al_3(BO_3)_4$ particles [53], the nonradiative relaxation with phonon emission from the $^4F_{3/2}$ state to the $^4I_{15/2}$ is overcome by thermal excitation from the $^4F_{3/2}$ level to upperlying ones, evidenced by the PL enhancement with the increase of the particles temperature. In other words, thermal quenching of the PL was not relevant here. For decorated particles (x = 0.20) the intensities at 750 nm (Figs. 5d,e) are similar to those under PA-like mechanism (*x* = 1.00) (Fig. 4c), but with lower $P_{exc}$.

## 4. Conclusion

We demonstrated that the efficiency of anti-Stokes excitation of trivalent neodymium ions ($Nd^{3+}$) in crystalline particles can be enhanced by decorating them with metallic nanoparticles (MNPs). Plasmonic effects were negligible due to the large distances between the MNPs and the $Nd^{3+}$, and the PL enhancement was attributed to the light-to-heat conversion in the interaction of the excitation laser light with the MNPs thanks to the excitation of vibrational transitions. Despite the light-to-heat conversion in the light interaction with MNPs is known, which can result in PL quenching, we demonstrated the PL enhancement due to the feasibility of the anti-Stokes excitation by phonon annihilation [43,51]. The introduction of MNPs into the dielectric particles could favor fluorescence quenching due to the energy transfer from the $Nd^{3+}$ to the MNPs. That undesirable effect is minimized when decorating the dielectric particles with MNPs due to the large distances among the MNPs and the $Nd^{3+}$. The results demonstrated here can be extended to other trivalent rare-earth ions, hosting media, MNPs, and excitation wavelengths.

**Acknowledgment**

We acknowledge financial support from the Brazilian Agencies: Fundação de Amparo à Pesquisa do Estado de São Paulo (FAPESP, Grants No. 2019/18828-1), Fundação de Amparo à Pesquisa do Estado de Alagoas (FAPEAL), Coordenação de Aperfeiçoamento de Pessoal de Nível Superior (CAPES) – Finance Code 001, Fundação de Amparo à Pesquisa do Estado de Goiás (FAPEG), Conselho Nacional de Desenvolvimento Científico e Tecnológico (CNPq), Financiadora de Estudos e Projetos (FINEP), scholarships in Research Productivity 2 under the Nr. 308753/2022-4, Nr. 303580/2021-6, and PhD scholarship (CNPq - 141759/2019-4), National Institute of Photonics (INCT de Fotônica). C. V. T.M., R. F. S. and D. F. L. thank CAPES for their Master scholarships. We also thank to J. G. B. Cavalcante for technical support. We thank Artur Bednarkiewicz for read our manuscript and provide some feedback.